# Observation of magnetism in Au thin films.

S. Reich*, G. Leitus and Y. Feldman.

Department of Materials and Interfaces,

The Weizmann Institute of Science, Rehovoth, Israel.

*e-mail: shimon.reich@weizmann.ac.il

Abstract:

**Direct magnetization measurements of thin gold films are presented. These measurements integrate the signal from the thin film under study and the magnetic contribution of the film's interface with the substrate. The diamagnetic contribution to the signal from the bulk substrate is of the same order as the noise level. We find that thin gold films can exhibit positive magnetization. The character of their magnetic behavior is strongly substrate dependent.**

In 1999 H.Hori et al. [1] observed magnetism in ~ 3 nm gold nanoparticles with an unexpected large magnetic moment of about 20 spins per particle. Since then several papers reported magnetism in gold nanoparticles with the emphasis on the stabilization of the particles by various polymers [2], size dependence of the magnetic moment [3] and observation of spin polarization of gold by x-ray magnetic circular dichroism [4]. The direct measurement of the magnetic properties of thin films is handicapped by the diamagnetic signal of the substrates. Thus, for example the magnetic signal from a 30 nm thin film composed of atoms exhibiting a considerable magnetization of 0.1 $\mu_B$ at 0.1 Tesla will be totally compensated by the diamagnetism of a quartz substrate 1 mm thick. The subtraction of background signal is a difficult and an inaccurate procedure.

In this paper an experimental method is presented which overcomes the problem of substrate subtraction.

A Quantum Design MPMS2 field- shielded magnetometer was used to measure magnetization. The sample in this instrument is dragged through superconducting pick up coils, which are wound in a second derivative arrangement – so called second derivative gradiometer [5]. This sensor will not produce output voltage for a uniform magnetic field or constant field gradient [6]. Therefore, if a uniform rod or pipe, which is considerably longer than the span of the gradiometer, is moved through the gradiometer pick up coils it will produce a constant magnetic flux, which will not contribute to the calculated dipolar magnetic moment. Such diamagnetic rods or pipes were used as substrates for the gold films in the experiments described below. Thin film gold rings were deposited on rods of borosilicate glass (accurate Pyrex capillaries) 15 cm long and 5.2 mm in diameter and on polypropylene drinking straws 17 cm long and 5 mm in diameter. The rings were deposited in vacuum ($5 \times 10^{-6}$ mbar) using an electron gun, and are ~0.6 cm high and 27 nm thick. The mass of the thin films is $5.2 \times 10^{-5}$ g, their surface area is ~ 1 cm$^2$.
The thin film rings are located ~25 mm from the center of these rods, see Fig. 1a. The gold used is 99.99 % pure.

As test samples for this experimental set up, 300 nm thick indium rings were deposited thermally from a boat onto Pyrex rods. Their thickness was chosen to be well above the percolation limit of In. These rings were split longitudinally so as not to trap magnetic flux in the superconducting state. The mass of these rings is ~ $2 \times 10^{-4}$ g. The raw data for magnetization vs. temperature and magnetic field, parallel to the surface of the film, in the supercoducting state are presented in Fig. 1b. In these zero field- cooled (ZFC) measurements there is no diamagnetic background.
All the critical parameters measured for In in the superconducing state are in accord with accepted values. $T_C$ = 3.4 K, $H_C$ (2K) =291 Oe The last value is larger than the bulk value at 2K: 184 Oe, due to finite size effects [7]. Note that in the normal state, T>$T_C$, the signal is close to zero without a diamagnetic contribution of the substrate.

In Fig. 2 the ZFC magnetization vs. field for gold films on glass and polypropylene are presented (raw data). Both samples show unusual magnetic character as compared to bulk gold. The gold film on polypropylene is weakly magnetic; ~ $3 \times 10^{-3}$ $\mu_B$ / atom at 1 Tesla. The magnetic response for these films is temperature independent, as the ZFC curves at 5K and at 300K practically coincide. The gold films on glass exhibit much stronger magnetization at 5K. However at 300K a very strong and linear diamagnetic response is observed. For comparison with the thin film data, we also show in Fig. 2 the magnetization measurement for a bulk gold sphere taken from the gold source used in the vacuum deposition process, the data are normalized to the mass of the thin films presented in the figure.

When a ZFC magnetization vs. temperature measurement at 0.1 Tesla for gold on Pyrex glass is performed, a classical paramagnetic response is obtained. This response is biased by a very strong diamagnetic contribution which is temperature independent, (see Fig.3). The data are well fitted by the Curie-Weiss law:

1) $\chi = \chi_0 + C/(T - \theta)$

$\chi_0 = -6.09 \times 10^{-3} \pm 2 \times 10^{-5}$ cm$^3$ / g-atom

$C = 0.102 \pm 1 \times 10^{-3}$ cm$^3$ × K / g-atom

$\theta = -0.6 \pm 2 \times 10^{-2}$ K

The $R^2$ value for the fit is 0.999.

The effective Bohr magneton number is: $\sqrt{(3k/N)} \times (1/\mu_B) \times \sqrt{C} = 0.90$

$\chi_0$ translated to volume susceptibility is $-6 \times 10^{-4}$, about 0.7% of the ideal diamagnetism, $-1/4\pi$.

It is expected that for thick enough gold films the measured molar magnetic susceptibility should be equal to that of bulk gold. A measurement for a 300 nm gold film on Pyrex is also presented in Fig. 3. The susceptibility observed is described very well by the Curie-Weiss law with an effective Bohr magneton number of 0.23, $\theta = -0.49$ K and $\chi_0 = -2 \times 10^{-4}$ cm$^3$/g-atom. We observe thus a sharp decrease in the effective magnetization as compared to that measured for the 27 nm film. At room temperature the measured diamagnetic susceptibility, $-1.6\times10^{-4}$ cm$^3$/g-atom, approaches from below the value for bulk gold, $-2.8\times10^{-5}$ cm$^3$/g-atom. Detailed study on the thickness dependence of the magnetization in thin gold films is in preparation.

We repeated the above measurements for five samples on polypropylene and six on Pyrex. The magnetic trends as function of field and temperature are reproducible. The maximum variation of the signal amplitude between samples on the same substrate is ~30%. We show in this letter results for samples with the best fit for dipole response curves, see inlay of Fig. 3 for the 27 nm film.

The level of magnetic impurities in the gold films has been measured by X-ray photoelectron spectroscopy (XPS):

Fe < 0.02 At.% ; Ni < 0.08 At.% ; Co < 0,04 At.% ; Mn < 0.04 At.%.

These impurities may contribute at 0.1 Tesla a signal in the range of $10^{-8}$ - $10^{-7}$ emu which is the limit of the sensitivity of our instrument. The negligible contribution of the magnetic impurities to the observed magnetization signal of the thin films is also apparent from the comparison between the magnetization curves for the thin films and that for bulk gold in Fig. 2. The blank substrates contribute a signal smaller than $10^{-6}$ emu in the temperature range: 2 K – 300 K and for fields up to 1 T. These blank measurements were performed by inverting the gold coated rods in the field so that a bare part of the rod ,blank substrate, was pulled through the gradiometer.

The nanometer level topography for the polypropylene and Pyrex substrates is presented in Fig. 4a. and 4b. These atomic force microscopy (AFM) pictures show that the Pyrex substrate is very smooth in comparison with polypropylene. This property of the substrates has a profound influence on texture of the vacuum- deposited thin gold films as revealed by X-ray diffraction (XRD). The film on Pyrex is (111) oriented and exhibits very good texture. The film on polypropylene shows a typical polycrystalline powder diffraction pattern, (see Fig.4c).

Bulk gold is diamagnetic: susceptibility = -2.8 x $10^{-5}$ $cm^3$ /g-atom. However its electronic configuration is a closed shell plus one s electron. For electronic configurations with an odd number of electrons the magnetic character is sometimes determined by the thermodynamic phase. Thus, for example, single atoms of alkali elements are magnetic in the gas phase. On the other hand alkali metal crystals and solid surfaces are not magnetic. This observation suggests that the magnetic character of a system may depend on its size and degree of condensation. Theoretically, magnetic moments have been found to be enhanced at surfaces and in thin films [8, 9]. In an interacting electron gas with a positive charge background in a solid, the magnetic character of the bulk is determined by the value of the Wigner radius, $r_s$, defined as the radius of the sphere occupied by the electron. Magnetic states are favorable for larger values of $r_s$. In an inhomogeneous system like a thin film, $r_s$ is a function of position $r_s$ (**r**). Lang and Kohn [10] have shown, that near a surface there is a dipole layer in which $r_s$ (**r**) is much larger than in the bulk. This may produce magnetism close to the surface of the metal. If this surface magnetization exhibits high enough anisotropy, it may explain the observed high value of magnetization at low temperature for gold thin films on the Pyrex substrate as these films are well textured. Gold on polypropylene is similar to Au nanoparticles. In this system the surface magnetization is averaged spatially over the sample.

The large temperature independent diamagnetism is apparently not Landau diamagnetism, a net magnetization anti- parallel to the external magnetic field, as it is about three orders of magnitude too large. However this diamagnetism is probably due to some very strong polarization at the interface between the gold and the glass substrate. We do not understand the mechanism of this polarization.

The experimental technique presented for the study of gold thin films is applicable to other metals where the magnetic character of the thin film and its interaction with the substrate are of interest.

Figure legends:

Fig. 1. (a): Thin films deposited on two substrates:

    I: 300 nm In on Pyrex capillary.

    II: 27 nm Au on Pyrex capillary.

    III: 27 nm Au on polypropylene.

  (b): Magnetization vs. temperature for an In 300 nm film measured in ZFC mode at 50 Oe. The magnetic field is parallel to the film.

  Inlay: Magnetization vs. field at 2K measured in the ZFC mode.

Fig. 2    Magnetization vs. magnetic field- ZFC mode -measured at 5K and 300K.

■ : Au, 27 nm film on Pyrex at 5K;

□ : Au, 27 nm film on Pyrex at 300 K;

▲ : Au, 27 nm film on polypropylene at 5K;

△ : Au, 27 nm film on polypropylene at 300 K;

○ : Au sphere (bulk gold, m = 273 mg) from the source used for the film deposition process, the values of magnetization are normalized to the weight of the gold thin films (m = $5.2 \times 10^{-2}$ mg). The gold sphere was measured inside a polypropylene straw from the same batch as those used for the deposition of the gold thin films.

Fig. 3.    Molar susceptibility vs. temperature-ZFC mode-

■ - for 27 nm Au on Pyrex measured at 0.1 T.

○ - for 300 nm Au on Pyrex measured at 0.5 T.

The line represents the best fit according to eq.1.

The inlay shows 290 magnetic dipole response curves, which compose the M vs. temp. curve for the 27 nm film.

Fig. 4   (a): AFM for the polypropylene substrate surface.

    (b): AFM for the Pyrex substrate surface.

    (c): XRD for Au on Pyrex and on polypropylene.

      Inlay: The (111) diffraction peak for Au film on polypropylene and on Pyrex.

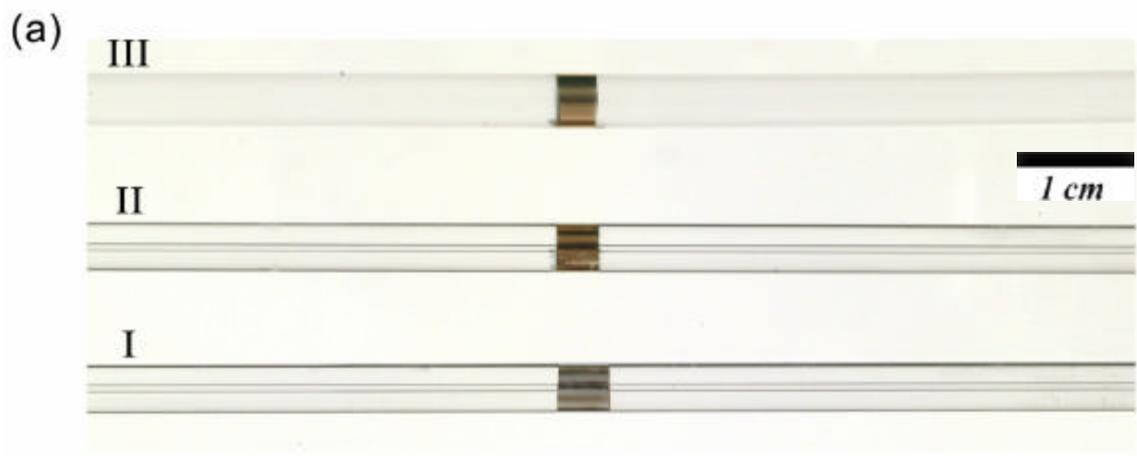
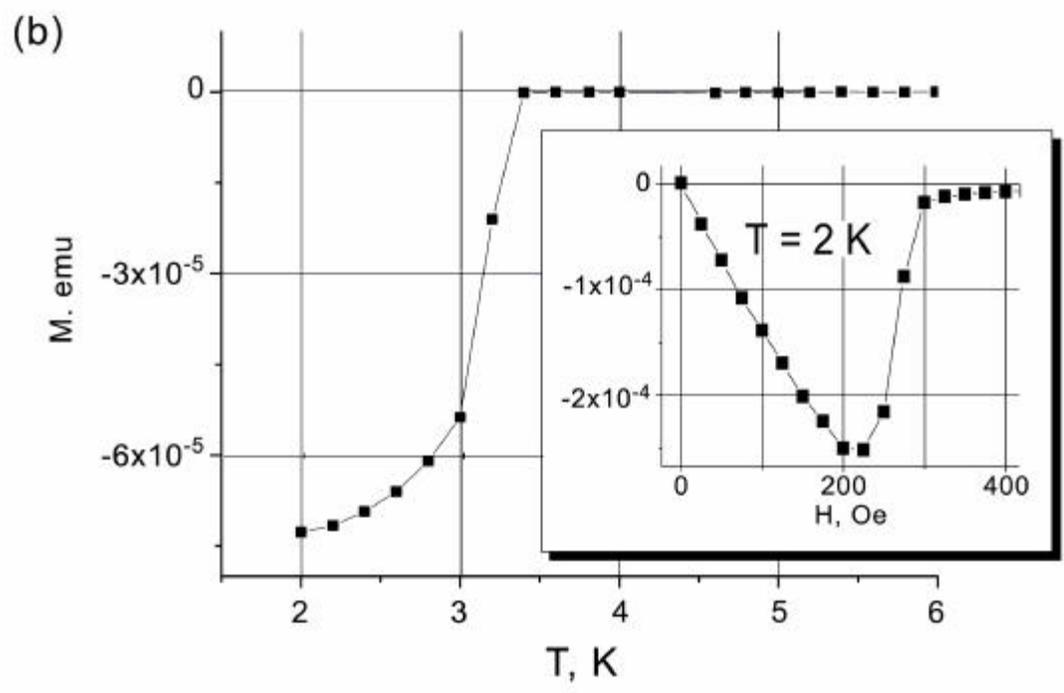

fig. 1

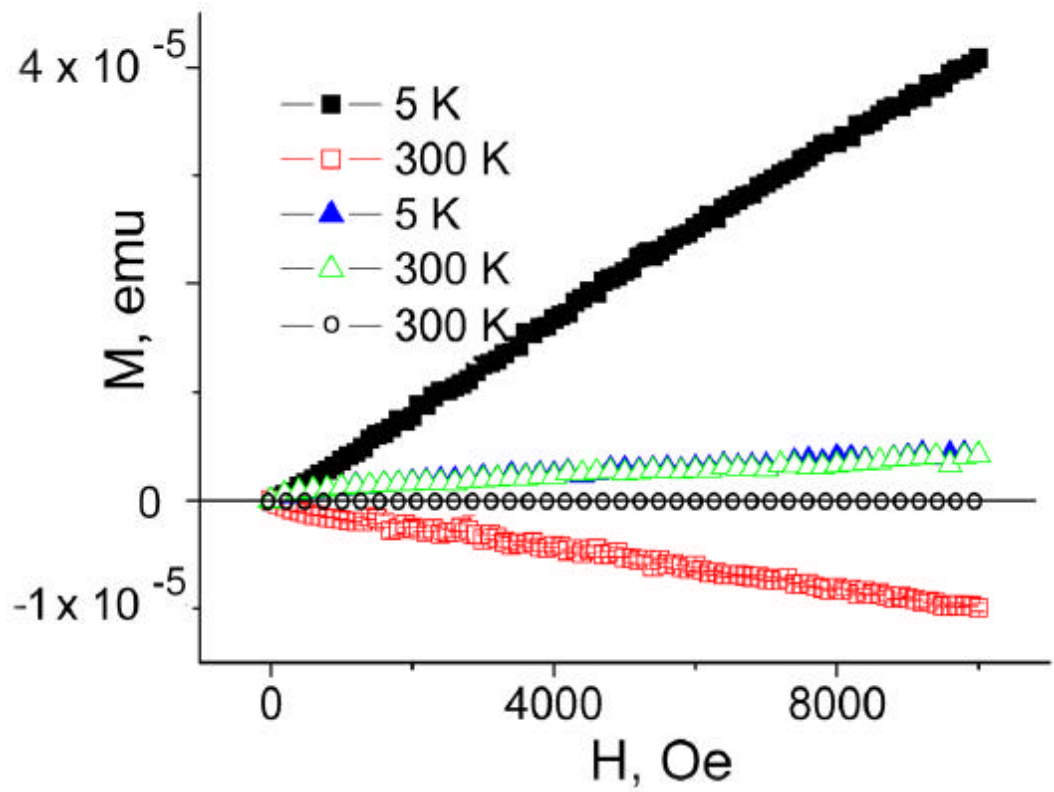

**fig. 2**

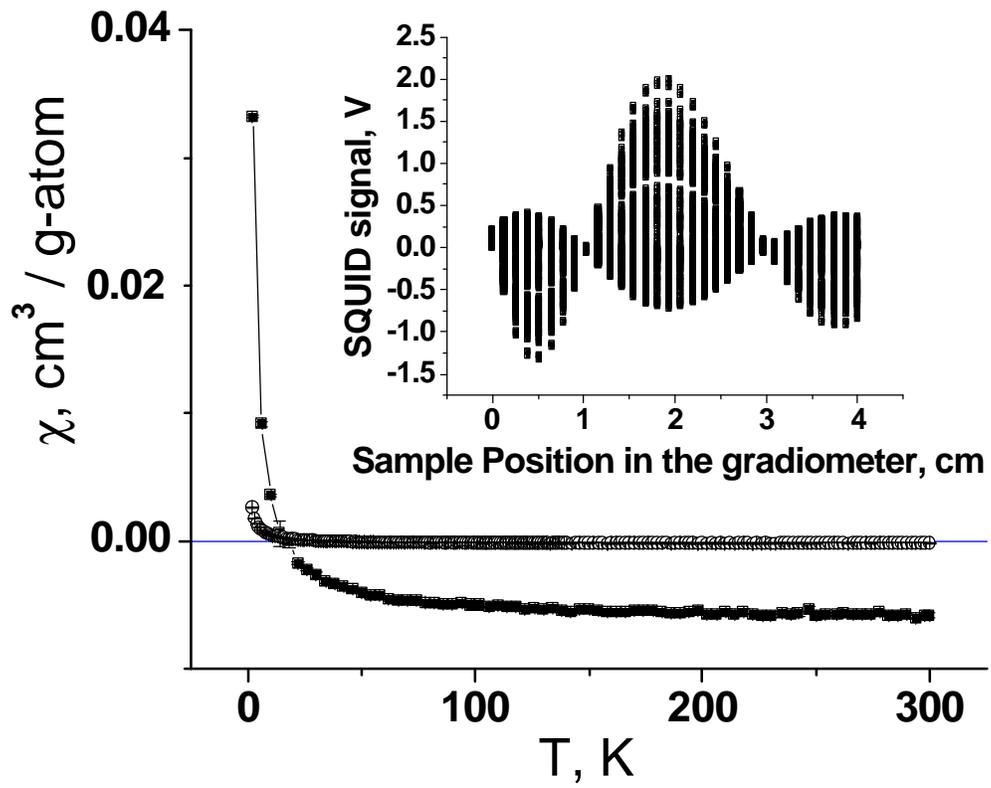

**fig. 3.**

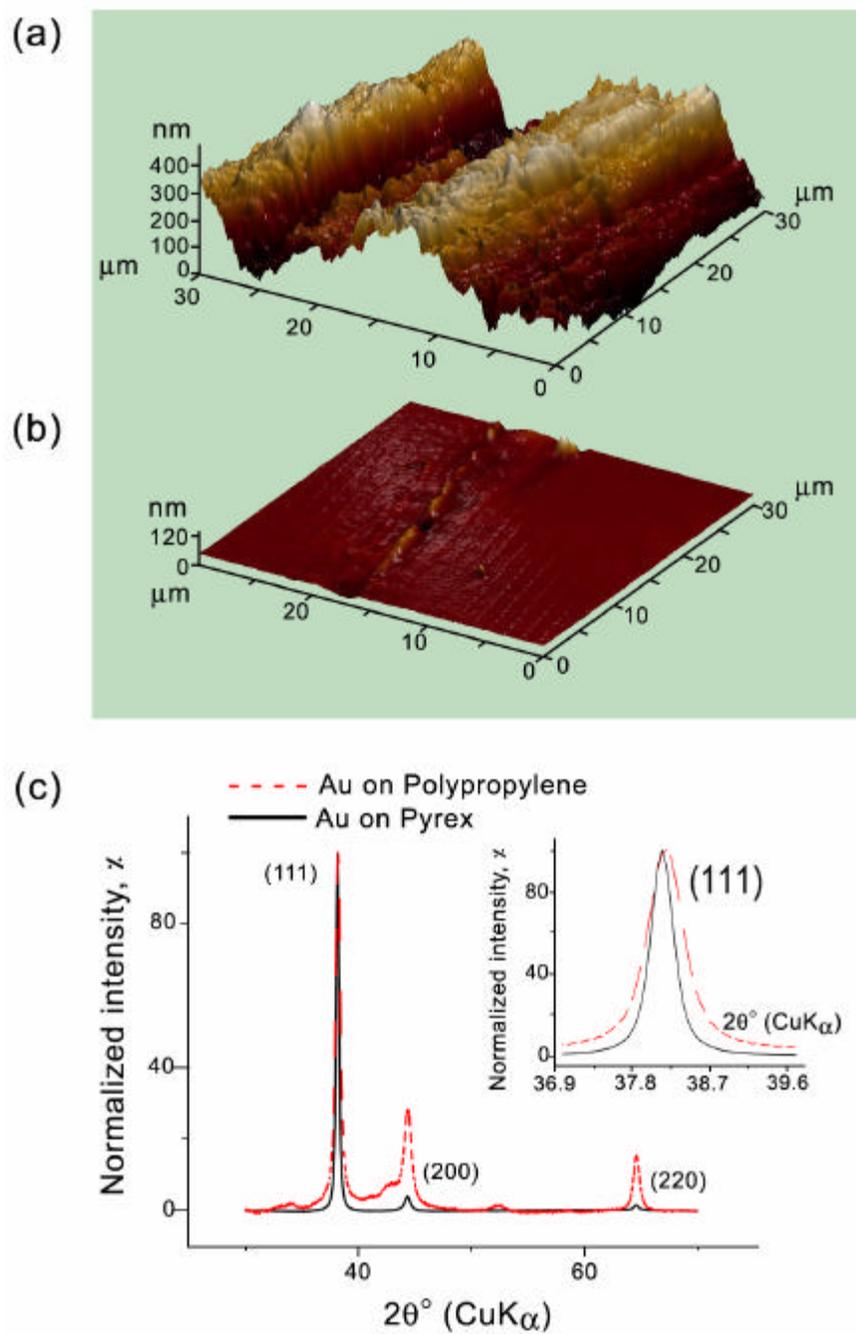

fig. 4